\documentclass[12pt,reqno]{article}
\usepackage[lmargin=2cm, rmargin=2cm]{geometry}
\usepackage{amsfonts}
\usepackage{amssymb}
\usepackage{enumerate}
\usepackage{amsthm}
\usepackage{array}
\usepackage{epsfig}
\usepackage[small,bf]{caption}
\usepackage{appendix}
\usepackage{dsfont}
\usepackage{hyperref}
\usepackage{amsmath}
\usepackage{graphicx}
\usepackage{epstopdf}
\usepackage{blkarray}
\usepackage{cite}

\allowdisplaybreaks \numberwithin{equation}{section}

\begin{document}

\begin{titlepage}
 \thispagestyle{empty}

\begin{flushright}
 \end{flushright}

 \begin{center}

 \vspace{30mm}

     { \LARGE{\bf  {Decoherence and Microscopic Diffusion at the Sachdev-Ye-Kitaev model}}}

     \vspace{40pt}

\Large{{\bf Javier M. Mag\'an}} \\[8mm]
{\small\slshape Instituto Balseiro, Centro At\'omico Bariloche \\
S.~C. de Bariloche, R\'io Negro, R8402AGP, Argentina\\

\vspace{5mm}

{\upshape\ttfamily javier.magan@cab.cnea.gov.ar}\\[3mm]}

\vspace{8mm}

     \vspace{10pt}

    \vspace{10pt}

\date{\today}

\end{center}

\begin{abstract}
Sachdev-Ye-Kitaev (SYK) or embedded random ensembles are models of $N$ fermions with random k-body interactions. They play an important role in understanding black hole dynamics, quantum chaos, and thermalization. We study out of equilibrium scenarios in these systems and show they display perfect decoherence at all times. This peculiar feature makes them very attractive in the context of the quantum-to-classical transition and the emergence of classical general relativity. Based on this feature and unitarity, we propose a rate/continuity equation for the dynamics of the $\mathcal{O}(e^N)$ microstates probabilities. The effective permutation symmetry of the models drastically reduces the number of variables, allowing for compact expressions of n-point correlation functions and entropy of the microscopic distribution. Further assuming a generalized Fermi golden rule allows finding analytic formulas for the kernel spectrum at finite $N$, providing a series of short and long time scales controlling the out of equilibrium dynamics of this model. This approach to chaos, long time scales, and $1/N$ corrections might be tested in future experiments.
\end{abstract}

 \vspace{10pt}
\noindent

\end{titlepage}

\thispagestyle{plain}

\tableofcontents

\baselineskip 6 mm

\newpage

\section{Black holes, decoherence and detailed balance}\label{secI}

Finding planckian corrections to black hole dynamics requires finding finite size corrections to thermalization processes, since the coupling $ \lambda \sim (G M^{2})^{-1}$, with $G$ Newton's constant and $M$ the black hole mass, is inversely proportional to black hole entropy $S_{\textrm{BH}}\sim 1/\lambda$. But finding finite size corrections to interacting many-particle states is an almost hopeless task, both analytically and numerically \cite{eisertreview,leareview}, a problem with various unwanted consequences. First, verifying unitarity, as predicted by AdS/CFT \cite{adscft}, by finding such corrections is bound to be a very difficult problem \cite{vijayreview,complehayden}. Second, this obstructs the exploration of time scales diverging with the entropy of the system (much larger than quasinormal relaxation \cite{quasireview}).

In this letter, we propose and explore one physical feature of black holes, namely dephasing/decoherence of the wavefunction, which has the potential to overcome part of these problems. In chaotic systems, we expect the appearance of three important time scales. A decoherence time scale $t_{\textrm{d}}$, the time at which reduced states are specified by its diagonal entries (the probabilities). A Markovian time scale $t_{\textrm{M}}$, the time at which the probabilities satisfy a Markovian equation. And a global relaxation time scale $t_{\textrm{r}}$, the time at which all diagonal elements saturate to their stationary values. It is plain that for any system $t_{\textrm{d}}\leq t_{\textrm{M}}\leq t_{\textrm{r}}$. The first inequality expresses that for times smaller than $t_{d}$, the state still contains important information in the off-diagonal entries. The second inequality notices that for times grater than $t_{\textrm{r}}$, the probabilities satisfy trivially the rate equation (see below). But the hierarchies between the three time scales might change from one system to another, and interesting simplifications might occur whenever $t_{\textrm{d}}\leq t_{\textrm{M}}\ll t_{\textrm{r}}$.

Since black holes are expected to saturate the bound on chaos \cite{bound}, we expect decoherence to appear in its strongest form. To show this is indeed the case, we will use Sachdev-Ye-Kitaev (SYK) type models \cite{kitaev,sachdev}:
\begin{equation}\label{SYK}
H=\sum\limits_{1\leq i_{1}<\cdots<i_{k}\leq N \atop 1\leq j_{1}<\cdots< j_{k}\leq N}J_{i_{1}\cdots i_{k};j_{1}\cdots j_{k}}c^{\dagger}_{i_{1}}\cdots c^{\dagger}_{i_{k}}c_{j_{1}}\cdots c_{j_{k}}\;,
\end{equation} 
where $J_{i_{1}\cdots i_{k};j_{1}\cdots j_{k}}$ are real independent random numbers with zero mean and variance equal to $J$, while $c^{\dagger}$ and $c$ create and destroy spinless fermions. In this version hermiticity requires $J_{i_{1}\cdots i_{k};j_{1}\cdots j_{k}}=J_{j_{1}\cdots j_{k};i_{1}\cdots i_{k}}$, each term contains $k$ annihilation and $k$ creation operators, and the number of particles $N_{T}=\sum\limits_{i=1}^{N}c^{\dagger}_{i}c_{i}$  is conserved \footnote{The author already worked \cite{usfree,usfreeblack} with this version in which $N_{T}=\sum\limits_{i=1}^{N}c^{\dagger}_{i}c_{i}$ is conserved. We remain in such formulation but techniqes apply to the Majorana case. The approach rests on the `all to all' random interactions and the permutation symmetry it implies.}. Average over the ensemble of $\mathcal{O}$ will be denoted by $\overline{\mathcal{O}}$.

These models were studied for quantum chaos \cite{kbody}. They were named k-body embedded ensembles and were proposed to improve the random Hamiltonian approximation \cite{haake}, which assumes interactions for every $k$. Recent interest grew mainly due to Kitaev \cite{kitaev}, who argued they have holographic duals and saturate the chaos bound \cite{bound}, and Sachdev \cite{sachdev}, who showed they have the right entropy. A complete discussion in these topics is given in \cite{remarks}, and interesting further aspects were developed in \cite{pol1, bilocal1,jensen,liouville,digital,bilocal2,local,general,higher1,garcia,susy, witten,sachdev2,Klebanov,strange,chethan}. Concurrently, these models were used to discuss aspects of eigenstate thermalization analytically \cite{ETH,deutsch,mark,usfree}, and a relation between large-N factorization and entanglement evolution was found \cite{usfreeblack}. Besides, these models might be constructed in the lab \cite{labSYK}. Finally, this model was used in \cite{chaosblack} to advance the program started by Maldacena in Ref. \cite{eternal}, followed by \cite{kleban,barbon1,simon,barbon2,echo}, concerning the quasiperiodicity of correlation functions in finite entropic systems.

We will work in Fock space, in the subspace of $m$ excited particles. The basis  $\vert i\rangle$ contains $i=1,\cdots ,\binom{N}{m}$ states, corresponding to all possible ways of locating the $m$ particles. Starting in one basis state $\vert i\rangle$, the global state at time $t$ reads:
\begin{equation}\label{rho}
\rho (t)=\sum\limits_{j,j'}\langle j\vert U(t)\vert i\rangle\langle i\vert U^{\dagger}(t)\vert j'\rangle\vert j\rangle \langle j'\vert\equiv\sum\limits_{j,j'}\rho_{jj'}(t)\vert j\rangle \langle j'\vert\;.
\end{equation}
To study decoherence in the evolution, first let us define $\langle j\vert H^{n}\vert i\rangle\equiv (H^{n})_{ji}$, and prove that
\begin{equation}\label{averageprod}
\overline{(H^{n})_{ji}(H^{m})_{ij'}}=
\sum\limits_{k_{1}k_{2}\cdots k_{n+m-2}}\overline{H_{jk_{1}}H_{k_{1}k_{2}}\cdots H_{k_{n-1}i}H_{ik_{n}}\cdots H_{k_{n+m-2}j'}}\propto \delta_{jj'}\;.
\end{equation}
This is proven as follows. For the Hamiltonian~(\ref{SYK}), a matrix element $H_{ij}$ is (if non-zero) one of the random couplings $J^{ij}$ associated to one term in the Hamiltonian $\hat{H}_{ij}$ that takes you from $j$ to $i$. Therefore, the right hand side of~(\ref{averageprod}) is a product of random couplings associated to each transition:
\begin{equation}\label{averageprod2}
\overline{(H^{n})_{ji}(H^{m})_{ij'}}=
\sum\limits_{k_{1}k_{2}\cdots k_{n+m-2}}\overline{J^{jk_{1}}J^{k_{1}k_{2}}\cdots J^{k_{n-1}i}J^{ik_{n}}\cdots J^{k_{n+m-2}j'}}\;.
\end{equation}
Since the random couplings have gaussian statistics, the average is the sum over all possible pairings of the couplings (Wick's theorem). Such pairings imply that if we used one term such as $\hat{H}_{k_{1}k_{2}}$ associated to one transtion from $k_{2}$ to $k_{1}$ with coupling $J^{k_{1}k_{2}}$, then the opposite transition $\hat{H}_{k_{2}k_{1}}$ (with the same coupling due to hermiticity) has to occur somewhere in the string~(\ref{averageprod}). This implies that for the average to be non zero, such string of operators associated to the string of random couplings takes the state $\langle j\vert$ in the left to the same state $\langle j\vert$, no matter the order in which contractions are made. Since this state has to be contracted at the end with $\vert j'\rangle$ we arrive at the right hand side of~(\ref{averageprod}). One can work out easily the cases with $n=m=1$ and $n=m=2$ to see in more detail the reasoning.

Since by expanding the unitary evolution in~(\ref{rho}), all terms are of the form~(\ref{averageprod}), we arrive at the following important conclusion:
\begin{equation}\label{pdiagin}
\overline{\rho (t)}= \sum\limits_{j}\overline{\vert\langle j\vert U( t)\vert i\rangle\vert^{2}}\, \vert j\rangle \langle j\vert\equiv\sum\limits_{j}p_{j}(t)\, \vert j\rangle \langle j\vert\;.
\end{equation}
For exactly the same reasons, dividing the system in $A$ and $B$ (any possible division), with Fock basis $j_{A}$ and $j_{B}$, provides diagonal reduced states on average:
\begin{eqnarray}
\overline{\rho_{A} (t)}&= &\overline{\textrm{Tr}_{B}\rho (t)} =\sum\limits_{j_{A}}p_{j_{A}}(t)\, \vert j_{A}\rangle \langle j_{A}\vert =\textrm{Tr}_{B}\overline{\rho (t)}\nonumber \\
\overline{\rho_{B} (t)}&= &\overline{\textrm{Tr}_{A}\rho (t)} =\sum\limits_{j_{B}}p_{j_{B}}(t)\, \vert j_{B}\rangle \langle j_{B}\vert =\textrm{Tr}_{A}\overline{\rho (t)}
\end{eqnarray}
We recognize an extremely peculiar feature of these models: starting at any fixed $\vert i\rangle$, the state is fully specified on average by its diagonal at any time. Global dephasing and local decoherence are as strong as they can be, the decoherence time being zero. One of the hierarquies is thus satisfied $t_{\textrm{d}}\ll t_{\textrm{r}}$. Notice that the same would happen in the generlized SYK models with spatial dimensions studied in \cite{local}.

Now, since off-diagonal entries are complex random numbers which vanish on average at all times, the `probability current' $p_{j}(t)$ is expected to satisfy a microscopic continuity equation:
\begin{equation}\label{rategeneral}
\frac{d p_{j}(t)}{dt}=\sum\limits_{i}\Gamma_{i\rightarrow j}(t)p_{i}(t)-(\sum\limits_{i}\Gamma_{j\rightarrow i}(t))p_{j}(t)\;,
\end{equation}
where $\Gamma_{i\rightarrow j}(t)$ are transition probability rates between microstates $i$ and $j$. We assume this is the case, i.e that the Markovian time scale is small as well, as it is assumed in other decoherence scenarios associated to different conserved currents. Unitarity implies the detailed balance condition:
\begin{equation}\label{detailed}
\sum\limits_{i}\Gamma_{i\rightarrow j}(t)=\sum\limits_{i}\Gamma_{j\rightarrow i}(t)\equiv \Gamma (t)\;,
\end{equation}
expressing the potential of unitarity to relate lifetimes to more complicated observables. Eq~(\ref{rategeneral}) is a novel perspective on large-N classicalization, containing $1/N$ effects and long time scales \footnote{In the context of generic large-N theories, such microscopic Fokker-Planck type equations have been rigorously justified by using large-N properties of coherent states, or the so-called de Finetti theorems \cite{usdef}.}.

One remark is needed. Notice that~(\ref{pdiagin}) transparently shows how the averaging procedure implies certain information loss and a natural coarse-graining (a usually unnoticed fact). It might seem strange we can recover $1/N$ effects after taking the average. A very non-trivial $1/N$ effect naturally taken into account by equations~(\ref{pdiagin}) and~(\ref{rategeneral}) is the following. Unitarity (in the form of detailed balance) ensures that the process drives an initial distribution towards the stationary state $p_{i}^{h}=1/\binom{N}{m}\sim \mathcal{O}(e^{-N})$, a strongly non-perturbative effect. This effect, missed in any gravitational description, was used in \cite{eternal} to discuss information loss in AdS/CFT. To the author knowledge, there are no dynamical computations obtaining these non-perturbative effects, only generic arguments involving unitarity. In the present framework, probabilities and correlation functions tend to non-zero values in a dynamical fashion. Given that at long times we reproduce these non-perturbative effects, it is natural to expect we recover other $1/N$ effects during evolution. At any rate, we remark that such result is the full answer on average, as proven by~(\ref{averageprod}).

\section{Exact consequences of permutation symmetry}\label{secII}

SYK displays an effective permutation symmetry on average, the relabeling of the fermions. In what follows, states differing from each other by moving a number $\alpha$ of particles are said to be distance $\alpha$ apart. 

Say $\vert \psi_{\textrm{in}}\rangle=\vert i\rangle$ (a given Fock basis state), the most important consequence of permutation symmetry is that the evolution of probabilities $p_{j}^{\alpha i}(t)$ of states $\vert j\rangle$ that are $\alpha$ particles apart from $\vert i\rangle$ is the same in all of them:
\begin{equation}
\vert \psi_{\textrm{in}}\rangle=\vert i\rangle\Rightarrow p_{j}^{\alpha i}(t)=p_{k}^{\alpha i}(t)\;.
\end{equation}
We only need to compute $p_{\alpha}\equiv p_{j}^{\alpha i}$ for $\alpha=0,\cdots ,m$, with $p_{\alpha=0}\equiv p_{i}$. Permutation symmetry reduces a seemingly exponentially hard problem of $\binom{N}{m}$ variables to a polynomially hard problem with $m+1\leq N/2+1$ variables. The task is to map~(\ref{rategeneral}) to $m+1$ coupled differential equations involving $p_{\alpha}$. To do so, notice the second consequence of permutation symmetry. The transitions rates $\Gamma_{i\rightarrow j}(t)\equiv \Gamma_{d}$ only depend on the distance $d$ between the states $i$ and $j$, and not on the states themselves.

Given these comments, the following result is exact provided~(\ref{rategeneral}) holds. The probability $p_{\alpha}(t)$ of states $\alpha$ particles away from $\vert i\rangle$, with $\vert \psi_{\textrm{in}}\rangle=\vert i\rangle$, satisfies:
\begin{equation}\label{klocal}
\frac{d p_{\alpha}}{dt}=-\Gamma (t) p_{\alpha} (t)+ \sum\limits_{\beta=0}^{m}C_{\alpha}^{\beta} (t) p_{\beta} (t)\;,
\end{equation}
where:
\begin{equation}\label{klocal2}
C_{\alpha}^{\beta}=\sum\limits_{j=0}^{2\alpha}\Gamma_{\beta-\alpha+j} (t) \sum \limits_{l=0}^{j}\binom{\alpha}{j-l}\binom{\alpha}{l}\binom{N-m-\alpha}{\beta-\alpha+j-l}\binom{m-\alpha}{\beta-\alpha+l}\;.
\end{equation}
The derivation was a case by case study. Here we explain its logic directly. The sum over $\beta$ adds contributions from states $\beta$ particles separated from $i$ to states $\alpha$ particles separated from $i$. We need to consider all distances $\beta$, explaining the sum in~(\ref{klocal}). For each $\beta$, we need to examine all possible distances $d$ from $\beta$ to $\alpha$. For distance $d$, the contribution to $\alpha$ reads:
\begin{equation}
\Gamma_{d}(t)\,\Omega_{d}\;,
\end{equation}
where $\Omega_{d}$ is the number of $\beta$ states $d$ away from $\alpha$, a number we need to compute. First, if the distance is not within $\beta-\alpha\leq d\leq\beta+\alpha $ the number of such contributions is zero. The reason goes as follows. Suppose that the $\alpha$ particles and $\alpha$ holes that differ from $\vert i\rangle$ in a state $\alpha$ are all located in different places than the $\beta$ particles and $\beta$ holes that differ from $\vert i\rangle$ in a state $\beta$. The distance between such states is $\beta+\alpha$, the maximum one. The minimum distance occurs when the $\alpha$ particles and $\alpha$ holes that differ from $\vert i\rangle$ in a state $\alpha$ are all contained in the $\beta$ particles and $\beta$ holes that differ from $\vert i\rangle$ in a state $\beta$. Here the distance is $\beta-\alpha$, the minimum one. Summing over $d$ is summing from $j=0$ to $j=2\alpha$ with $d_{j}=\beta-\alpha+j$, explaining the first sum in~(\ref{klocal2}), where one should notice that $2\alpha -j$ is the number of particles plus holes shared by states $\alpha$ and $\beta$ but not shared with $\vert i\rangle$. This $2\alpha -j$ sharing can occur in various ways. They can share $\alpha$ particles and $\alpha-j$ holes, or $\alpha-1$ particles and $\alpha-j+1$ holes...or $\alpha-j-l$ particles and $\alpha-j+l$ holes...until sharing $\alpha-j$ particles and $\alpha$ holes. The number of each possibility is a straightforward balls and bins problem which justifies the last sum in $l$ in eq~(\ref{klocal2}).

The next results concerns entropy and correlation functions. They do not depend on the Markovian approximation. They are exact on average given~(\ref{pdiagin}). Such equation implies that the obervables of direct interest are of the form:
\begin{equation}\label{diagonal}
\mathcal{O}=\sum\limits_{k}\mathcal{O}_{k}\vert k\rangle\langle k\vert\;,
\end{equation}
where $k$ runs over all Fock microstates. Any string as:
\begin{equation}
\mathcal{O}_{ij\cdots l}=n_{i}\, n_{j}\,\cdots \, n_{l}\;,
\end{equation}
where $n_{i}\equiv c^{\dagger}_{i}c_{i}$, belongs to the diagonal algebra~(\ref{diagonal}). Taking again $\vert i\rangle$, with the first $m$ particles excited as the initial state, and setting $j\leq m$ and $j'\leq N-m$, the generic n-point correlation function is:
\small
\begin{equation}\label{exp}
\langle n_{i}\cdots n_{j}n_{m+1}\cdots n_{m+j'}\rangle =\sum\limits_{\alpha=j'}^{m-j}\binom{m-j}{\alpha}\binom{N-m-j'}{\alpha-j'}p_{\alpha}(t)\;.
\end{equation}
\normalsize
The product of binomials is the number of states at distance $\alpha$ giving non-zero contribution (equal to one). Thus computing $p_{\alpha}(t)$ allows finding all n-point correlations, without worrying about the number of insertions.

Finally, entanglement entropy of the diagonal algebra reads:
\begin{eqnarray}
S(\rho_{D})&=&-\sum\limits_{k=1}^{\binom{N}{m}}p_{k}(t)\log p_{k}(t)= \nonumber \\ &=&-\sum\limits_{\alpha=0}^{m}\binom{m}{\alpha}\binom{N-m}{\alpha}p_{\alpha}(t)\log p_{\alpha}(t)\;.
\end{eqnarray}
Given that the saturation of this entropy implies saturation of all operators belonging to the diagonal algebra, one is tempted to conclude it relates to complexity evolution \cite{com1,com2,pepe1,pepe2,roberts,shira,myers}. Care has to be taken since as we show below this entropy saturates at a time of order $t\sim\mathcal{O}(N)$, much faster than expected for complexity.

\section{One dimensional k-local diffusion}\label{secIII}

The probabilities $p_{\alpha}(t)$ are on average a convenient and sufficient set. To proceed we need an expression for the transition rates. In this section we develop two approximations of increasing strength. First, as in other continuity equations, we can assume the transition rates to be non zero only between states directly connected by the Hamiltonian \footnote{This assumption suggests considering a `Hamiltonian graph'. The vertices $V=\lbrace\vert i\rangle;i=1,\cdots , \binom {N}{m}\rbrace$ are the Fock basis, and the edges are the transition rates $E_{ij}=\Gamma_{d_{ij}}$. A characteristic aspect of SYK is that the graph is regular, homogeneous, isotropic and weighted, with valency $v_{k}=\sum\limits_{\alpha=1}^{k}\binom{m}{\alpha}\binom{N-m}{\alpha}$. It is \emph{almost} an expander graph \cite{avi1,avi2,us3}, since the diameter is $D\sim \log \Omega$, with $\Omega$ the number of vertices in the graph. The rate equation~(\ref{klocalap1}) is a random walk on this graph.}. This natural `locality' assumption, expected to be accurate in the thermodynamic limit, results in the following k-local diffusion equation:
\begin{equation}\label{klocalap1}
\frac{d p_{\alpha}}{dt}=-\Gamma (t) p_{\alpha}+\sum\limits_{\beta=\alpha-k}^{\alpha+k}C_{\alpha}^{\beta} (t) p_{\beta}\;,
\end{equation}
where $C_{\alpha}^{\beta} (t)$ is defined as above. In this approximation we only need to measure/compute $\Gamma_{1},\cdots \Gamma_{k}$, two functions for the physical $k=2$ case with two-body interactions. Then~(\ref{klocalap1}) together with the results of the previous section would provide the correlation functions for the non-equilibrium process.

A stronger approximation is the following:
\begin{equation}\label{assumption}
\Gamma_{i\rightarrow j}(t)=D(t)\overline{\vert\langle j\vert H\vert i\rangle\vert^{2}}\;,
\end{equation}
for some function $D (t)$. Notice that at small times $D (t)\propto t$ for any Hamiltonian, while Fermi golden rule $D (t)= D $ is expected to hold after some time and last for longer times, given that in the thermodynamic limit each basis state decays to a continuum in SYK. These regimes have been observed in SYK models \cite{leareview}. The Hamiltonian matrix element between $\vert i\rangle$ and $\vert j_{\alpha}\rangle$ separated by distance $\alpha$ is:
\begin{equation}\label{averageH}
\overline{\vert\langle j_{\alpha}\vert H\vert i\rangle\vert^{2}}=\overline{(\sum\limits_{\textrm{paths}} J_{\textrm{paths}})^{2}}= \Omega_{\textrm{paths}}J^{2}=\binom {m-\alpha}{k-\alpha} J^{2}\;,
\end{equation}
since after moving $\alpha$ particles we have $\binom {m-\alpha}{k-\alpha}$ `paths' to create and destroy fermions without affecting the final state. Since each path contributes a random number, averaging the paths sum squared gives the number of paths times the variance of one path.
With this approximation we arrive at:
\begin{equation}\label{kap}
\frac{d p_{\alpha}}{dt}=-\Gamma (t) p_{\alpha}+D (t)\sum\limits_{\beta=\alpha-k}^{\alpha+k}C_{\alpha}^{\beta}p_{\beta}\;,
\end{equation}
where:
\begin{equation}\label{kap2}
C_{\alpha}^{\beta}=\sum\limits_{j=0}^{2\alpha}\binom{m-(\beta-\alpha+j)}{k-(\beta-\alpha+j)} \sum \limits_{l=0}^{j}\binom{\alpha}{j-l}\binom{\alpha}{l}\binom{N-m-\alpha}{\beta-\alpha+j-l}\binom{m-\alpha}{\beta-\alpha+l}\;.
\end{equation}
Notice that detailed balance still holds:
\begin{equation}
\sum\limits_{\beta=\alpha-k}^{\alpha+k}C_{\alpha}^{\beta}=\sum\limits_{\alpha=0}^{\alpha=k}\binom{m-\alpha}{k-\alpha}\binom{m}{\alpha}\binom{N-m}{\alpha}=\frac{\Gamma (t)}{D(t)}\;.
\end{equation}
The interesting aspect of this approximation is that it can almost be solved exactly, obtaining non-trivial aspects of the long time dynamics, as we show in the next section.

\section{Kernel spectrum and time scales}\label{secIV}

The diffusion process can be written as:
\begin{equation}
\frac{dp}{dt}=\Gamma (t) Mp(t)\;,
\end{equation}
where the $(m+1)\times (m+1)$ matrix $M$ is defined to match eqs~(\ref{kap}) and~(\ref{kap2}). The solution is:
\begin{equation}\label{eqap}
p(t)=\sum\limits_{a=1}^{m+1}c_{a}e^{\lambda_{a}\int\limits_{0}^{t}\Gamma (t')dt'}v^{a}\;,
\end{equation}
where $\lambda_{a}$ and $v^{a}$ are the $m+1$ eigenvalues and eigenvectors of $M$, and the $c_{a}$ are fixed by the initial condition $p_{\alpha}(t)=\delta_{\alpha,0}$.  As usual, $c_{1}v_{1}=1/\binom{N}{m}$ and $\lambda _{1}=0$, corresponding to the stationary distribution. In what follows we set $N=2m$.

The equation can be solved by numerical methods. When doing so we found something striking and unexpected. The kernel spectrum is composed by series of fractional numbers following specific laws \footnote{We are indebted to Eduardo Test\'e for collaborating in finding the underlying patern of the kernel spectrum.}. We could not prove analytically the following relations, but the reader can verify them for any $N=2m$. The kernel spectrum $\lambda_{i}^{k}(m)$, organized as $\lambda_{1}^{k}(m)=0>\lambda_{2}^{k}(m)\geqslant\cdots\geqslant \lambda_{m+1}^{k} (m)=-1$, reads:
\begin{equation}\label{genericspec}
\lambda_{i}^{k}(m)=\frac{(i-1)(i-2(1+m))(\sum\limits_{q=0}^{2k-2}P(m,q)i^{q})}{(m-k+1)\cdots (m-1)m(m+1)\cdots (m+k)}\;,
\end{equation}
where $P(m,2k-2)=1$ and $P(m,0)=k(m-k+2)\cdots m(m+2)\cdots (m+k)$. For $k=1,2$
\begin{eqnarray}\label{k12}
\lambda_{i}^{1}(m)&=&\frac{(i-1)(i-2(1+m))}{m(m+1)} \nonumber\\
\lambda_{i}^{2}(m)&=&\frac{(i-1)(i-2(1+m))(i^{2}-i(3+2m)+2m(2+m))}{(m-1)m(m+1)(m+2)} \;.
\end{eqnarray}
Formulas for $k=3,4,5$ were explicitly computed but a generic expression as a function of $k$ was not found.

Eqs~(\ref{genericspec}) and~(\ref{k12}) are the last main results of the letter. They provide $m$ time scales which might be measured in experiments \cite{labSYK} (more precisely one can measure all ratios). It is remarkable such a formula for the $i$-eigenvalue even exists, suggesting the influence of some hidden symmetry \footnote{This is further suggested by the structure of eigenvectors, basically equal for all $k$.}.

We also remark that the second eigenvalue (the gap), which controls the late time approach to equilibrium:
\begin{equation}\label{larget}
p_{\alpha}(t\gg t_{\textrm{r}})=\frac{1}{\binom{N}{m}}+c_{2}e^{\lambda_{2}^{k}(m)\int\limits_{0}^{t}\Gamma (t')dt'}v^{2}_{\alpha}\;,
\end{equation}
and defines the global relaxation time $t_{\textrm{r}}$ (the time where probabilities saturate to their stationary value), is found to be:
\small
\begin{equation}\label{gap}
c_{2}(m)=\frac{1-2m}{\binom{N}{m}}~~~~~~~~~\lambda_{2}^{k}(m)=-\frac{2k}{m+k}~~~~~~~~~v^{2}_{\alpha}=\frac{-m+2\alpha}{m}\;.
\end{equation}
\normalsize
For $k\sim \mathcal{O}(1)$ we see the natural appearance of time scales of $\mathcal{O}(N)$ (in units of $\tau=\int\limits_{0}^{t}\Gamma (t')dt'$) at the edge of the spectrum, the longest time scales controlling the out of equilibrium process. We conclude that $t_{\textrm{r}}/\Gamma\sim \mathcal{O}(N)$, the kernel gap time scale.

Combining this result with the expression for the correlators~(\ref{exp}), a definite prediction is that at long times, all correlation functions decay with the same exponent, albeit with different prefactors and to different values.

\section{Conclusions}\label{secV}


Black hole dynamics is expected to be maximally chaotic \cite{bound}. As such, we expect such systems to display the strongest forms of decoherence. This is an interesting aspect of black hole dynamics which has not be considered in the literature.

In this letter, we explored SYK, dual to quantum gravity in AdS2 \cite{kitaev,remarks}, in non-equilibrium scenarios from this perspective. We showed that on average, the state (and reduced states) displays exact dephasing/decoherence at all times, since off-diagonal terms vanish~(\ref{pdiagin}). Given such feature, and reminding that decoherence roots classical diffusion/continuity equations for conserved currents, we analyzed the generic rate equation~(\ref{rategeneral}) for the conserved current at hand, formed by the average microscopic probabilities of the Fock space basis. Such equation interconnects all equal time correlation functions of the theory.  We then analyzed aspects of symmetry, correlation functions and time scales. We conclude that SYK models are an interesting and convenient environment to study the quantum-to-classical transition, since decoherence can be discussed analytically.

From a paralell perspective, equations~(\ref{pdiagin}) and~(\ref{rategeneral}) are an interesting new look to large-N classicalization, which we expect to generalize to other microscopic models of black holes. This indeed has been shown in \cite{usdef}, by showing strong decoherence in the coherent state basis of large-N theories, which implies that any reduced dynamics should be conveniently described by Fokker-Planck type equations in the so-called coherence group. The generic lessons are thus the following. Due to strong decoherence in the coherent state basis, and that microstates decay to a continuum in the thermodynamic limit, the diagonal algebra in such basis is controlled by a rate equation (like a nuclear chain reaction process) typical of an open system, with the bath being the non-diagonal operator algebra. The entanglement between both algebras is the entropy of the probability distribution, and at long times entanglement saturates to the thermal (black hole) entropy. Such rate equation contains $1/N$ effects, long time scales, and controls many non-trivial correlation functions, including those with a number of insertions scaling with the total number of degrees of freedom.

\section*{Acknowledgements}

It is a pleasure to thank Eduardo Test\'e Lino for helping in the derivation of the kernel spectrum. We also wish to thank Horacio Casini for discussions around the diagonal operator algebra, and Juan Maldacena and Jos\'e Barb\'on for comments on the draft.
This work was supported by the Simons foundation through the It From Qubit Simons collaboration.





\newpage
\begin{thebibliography}{00}

   \bibitem{eisertreview}  
 C.~Gogolin and J.~Eisert,
 {\em Equilibration, thermalisation, and the emrgence of statistical mechanics in closed quantum systems},
 Rep.\ Prog.\ Phys. {\bf 79} (2016) 5,
 [arXiv:1503.07538 [quant-ph]].

\bibitem{leareview}
 F.~Borgonovi, F.~M.~Izrailev, L.~.F~Santos and V.~G.~Zelevinsky,
{\em Quantum chaos and thermalization in isolated systems of interacting particles},
Phys.\ Rep. {\bf 626} (2016) 1-58,
  [arXiv:1602.01874 [cond-mat]].
  
      \bibitem{adscft}
  J.~M.~Maldacena,
  {\em The Large N limit of superconformal field theories and supergravity},
  Adv.\ Theor.\ Math.\ Phys {\bf 2} (1998),
  [arXiv:9711200 [hep-th]].
  
   \bibitem{vijayreview}
  V.~Balasubramanian and B.~Czech,
  {\em Quantitative approaches to information recovery from black holes},
  Class.\ Quant.\ Grav.\  {\bf 28} (2011) 163001,
  [arXiv:1102.3566 [hep-th]].
  
  
      \bibitem{complehayden}
  D.~Harlow and P.~Hayden,
  {\em Computational complexity vs firewalls},
  JHEP {\bf 1306} (2013) 085,
  [arXiv:1301.4504 [hep-th]].
  
  
 
  \bibitem{quasireview}
E.~Berti, V.~Cardoso and A.~O.~Starinets, 
 Class.\ Quant.\ Grav {\bf 26} (2009) 163001,  
[arXiv:0905.2975 [hep-th]].

  \bibitem{kitaev}
  A.~Kitaev,
  {\em A simple model of quantum holography},
  Talks at KITP, April 7, 2015 and May 27, 2015.


   \bibitem{sachdev}
   S.~Sachdev and J.~Ye,
   {\em Gapless spin fluid ground state in a random, quantum Heisenberg ferromagnet},
   Phys. \ Rev.\ Lett.\ {\bf 70} (1993) 3339,
   arXiv:cond-mat/9212030 [cond-mat]. 
   
   
  S.~Sachdev,
  {\em Bekenstein-Hawking Entropy and Strange Metals},
  Phys.\ Rev.\ X {\bf 5} (2015) 041025,
  [arXiv:1506.05111 [hep-th]].
  

  
 \bibitem{usfree}
  J.~M.~Magan,
  {\em Random free fermions: An analytical example of eigenstate thermalization},
  Phys.\ Rev.\ Lett {\bf 116} (2016) 030401, 
  [arXiv:1508.05339 [quant-ph]].
   
       \bibitem{usfreeblack}
  
 J.~M.~Magan,
{\em Black holes as random particles: entanglement evolution in infinite range and matrix models},
 JHEP {\bf 1608} (2016) 081,
 [arXiv:1601.04663 [hep-th]].
 
 \bibitem{kbody}
   L.~Benet and H.~A.~Weidenmueller,
  {\em Review of the k-body embedded ensebmles of gaussian random matrices},
 J.\ Phys.\ A {\bf 36} (2003) 3569,
  [arXiv:0207656 [cond-mat]].

      \bibitem{haake}
F.~Haake,
{\em Quantum Signatures of Chaos},
Springer-Verlag Berlin Heidelberg, {\bf 54} (2010).  


   \bibitem{bound}
J.~Maldacena, S.~H,~Shenker and D.~Stanford,
{\em A bound on chaos},
JHEP {\bf 08} (2016) 106,
  [arXiv:1503.01409 [hep-th]].
  



    \bibitem{remarks}
J.~Maldacena and D.~Stanford,
{\em Remarks on the Sachdev-Ye-Kitaev model},
Phys.\ Rev. \ D.\ {\bf 94} (2016) 106002.
  
  
 

 
 
 
      \bibitem{pol1}
J.~Polchinski and V.~Rosenhaus,
{\em The spectrum in the Sachdev-Ye-Kitaev model},
JHEP {\bf 161} (2016),
  [arXiv:16001.06768 [hep-th]].
  
   \bibitem{bilocal1}
  A.~Jevicki, K.~Suzuki and J.~Yoon,
  {\em Bilocal holography in the SYK model},
 JHEP {\em 1607} 2016 007,
  [arXiv:1603.06246 [hep-th]].
  
  
      \bibitem{jensen}
K.~Jensen,
{\em Chaos in AdS2 holography},
Phys.\ Rev. \ Lett.\ {\bf 117} (2016) 111601,
   [arXiv:1605.06098 [hep-th]].
 
 
    \bibitem{liouville}
D.~Bagrets, A.~Altland and A.~Kamenev,
{\em Sachdev-Ye-Kitaev model as Liouville quantum mechanics},
Nuc.\ Phys.\ B {\bf 911} (2016) 191-205,
[arXiv:1607.00694 [cond-mat]].
 
 \bibitem{digital}
L.~Garcia-Alvarez, I.~L.~Egusquiza, L.~Lamata, A.~del Campo, J.~Sonner and E.~Solano,
{\em Digital quantum simulation of minimal AdS/CFT},
  [arXiv:1607.08560 [hep-th]].  

   \bibitem{bilocal2}
A.~Jevicki and K.~Suzuki,
{\em Bi-local holography in the SYK model: perturbations},
  [arXiv:1608.07567 [hep-th]].
  
 


      \bibitem{local}
Y.~Gu, X.~Qi and D.~Stanford,
{\em Local criticality, diffusion and chaos in generalized Sachdev-Ye-Kitaev models},
[arXiv:1609.07832 [hep-th]].



        \bibitem{general}
D.~Gross and V.~Rosenhaus,
{\em A generalization of Sachdev-Ye-Kitaev},
[arXiv:1610.01569 [hep-th]].
  
    \bibitem{higher1}
M.~Berkooz, P.~Narayan, M.~Rozalli and J.~Simon,
{\em Higher dimensional generalizations of the SYK model},
  [arXiv:1610.02422 [hep-th]].    
  
      \bibitem{garcia}
A.~M.~Garc\'ia -Garc\'ia and J.~J.~M.~Verbaarschot,
{\em Spectral and thermodynamics properties of the Sachdev-Ye-Kitaev model},  
   [arXiv:1610.03816 [hep-th]].
   
    
       \bibitem{susy}
W.~Fu, D.~Gaiotto, J.~Maldacena and S.~Sachdev,
{\em Supersymmetric SYK models},
  [arXiv:1610.08917 [hep-th]].
  
  

  
      \bibitem{witten}
E.~Witten,
{\em An SYK-like model without disorder},  
   [arXiv:1610.09758 [hep-th]].

     \bibitem{sachdev2}
S.~Sachdev and A.~A.~Patel,
{\em Quantum chaos on a critical Fermi surface},
 [arXiv:1611.00003 [cond-mat]].  
  
 
  
  \bibitem{Klebanov}
I.~R.~Klebanov, G.~ Tarnopolsky,
{\em Uncolored random tensors, melon diagrams, and the SYK models},  
   [arXiv:1611.08915 [hep-th]].
   
    \bibitem{strange}
R.~A.~Davison, W.~Fu, A.~Georges, Y.~Gu, K.~Jensen and S.~Sachdev,
{\em Thermoelectric transport in disordered metals without quasiparticles: the SYK models and holography},  
   [arXiv:1612.00849 [cond-mat]].
   
   
   
   \bibitem{chethan}
C.~Krishnan, S.~Sanyal and P.~N.~Balasubramanian,
{\em Quantum chaos and holographic tensor models},  
   [arXiv:1612.06330 [hep-th]].   
 
 
 
 
 
   
\bibitem{ETH}
  R.~V.~Jensen and R.~Shankar,
  {\em Statistical behavior in deterministic quantum systems with few degrees of freedom},
  Phys.\ Rev.\ Lett {\bf 54} (1985) 1879.

\bibitem{deutsch}
J.~M.~Deutsch, 
{\em Quantum statistical mechanics in a closed system},
Phys.\ Rev.\ A {\bf 43} (1991) 2046.

\bibitem{mark}
M.~Srednicki,
{\em Chaos and quantum thermalization},
Phys.\ Rev.\ E {\bf 50} (1994) 888,
  [arXiv:9403051 [cond-mat]].
  


 
  \bibitem{labSYK}
  
 I.~Danshita, M.~Hanada and M.~Tezuka,
{\em Creating and probing the Sachdev-Ye-Kitaev model with ultracold gases: Towards experimental studies of quantum gravity},
 [arXiv:1606.02454 [cond-mat]].
 

 

    \bibitem{chaosblack}
J.~S.~Cotler,G.~Gur-Ari, M.~Hanada, J.~Polchinski, P.~Saad, S.~Shenker, D.~Stanford, A.~Streicher, M.~Tezuka,
 {\em Black holes and random matrices},
[arXiv:1611.04650 [hep-th]].
 

 \bibitem{eternal}
J.~M.~Maldacena,
  {\em Eternal black holes in AdS},
  JHEP {\bf 0304}, 021 (2003),
  [arXiv:0106112 [hep-th]].
  

  
 
\bibitem{barbon1}
J.~L.~Barbon and E.~Rabinovici
{\em Very long time scales and black hole thermal equilibrium}
JHEP {\bf 0311}, 047 (2003)
 [arXiv:0308063 [hep-th]].
 
 \bibitem{kleban}

M.~Kleban, M.~Porrati and R.~Rabadan,
{\em Poincare recurrences and topological diversity},
  JHEP {\bf 0410}, 030 (2004),
  [arXiv:hep-th/0407192].
  
   \bibitem{simon}
V.~Balasubramanian, S.~Ross, J.~Simon
{\em Black holes, entanglement and random matrices}
Class.\ Quant.\ Grav {\bf 31} (2014) 18 
 [arXiv:1404.6198 [hep-th]].
 
 

 \bibitem{barbon2}
J.~L.~Barbon and E.~Rabinovici
{\em Geometry and quantum noise}
Fortschr.\ Phys {\bf 62} (2014) 8 
 [arXiv:1404.7085 [hep-th]].
 
 
 \bibitem{echo}
V.~Balasubramanian, B.~Craps, B.~Czech and G.~S\'arosi ,
{\em Echoes of chaos from string theory black holes},
 [arXiv:1612.04334 [hep-th]].
 

 
  \bibitem{usdef}
J.~M.~Magan,
{\em De Finetti theorems and entanglement in large-N theories and gravity},
 [arXiv:1705.1885072 [hep-th]].

 

 
  
  \bibitem{com1}
  A.~.R~Brown, D.~A.~Roberts, L.~Susskind, B.~Swingle and Y.~Zhao,
  {\em Complexity equals action},
  Phys.\ Rev. \ Lett. {\bf 116} (2016) 191301,
  [arXiv:1509.07876 [hep-th]].
 
\bibitem{com2}
  A.~.R~Brown, D.~A.~Roberts, L.~Susskind, B.~Swingle and Y.~Zhao,
  {\em Complexity, action, and black holes},
  Phys.\ Rev. \ D. {\bf 93} (2016) 086006,
  [arXiv:1512.04993 [hep-th]].
   
 
    \bibitem{pepe1}
  J.~L.~F.~Barb\'on and E.~Rabinovici,
  {\em Holographic complexity and cosmological singularities},
  JHEP {\bf 1601} (2016) 084,
  [arXiv:1509.09291 [hep-th]].

     \bibitem{pepe2}
  J.~L.~F.~Barb\'on and J.~Martin-Garc\'ia ,
  {\em Holographic complexity of cold hyperbolic black holes},
  JHEP {\bf 1511} (2015) 181,
  [arXiv:1510.00349 [hep-th]].
  
     \bibitem{roberts}
   D.~A.~Roberts and B.~Yoshida
{\em Chaos and complexity by design},
  [arXiv:1610.04903 [hep-th]]. 
 
   \bibitem{shira}
  S.~Chapman, H.~Marrochio and R.~C.~Myers,
  {\em Complexity of formation in holography},
  [arXiv:1610.08063 [hep-th]].
  
 
  
 \bibitem{myers}
  D.~Carmi, R.~C.~Myers and P.~Rath,
  {\em Comments on holographic complexity},
  [arXiv:1612.00433 [hep-th]].
  
     \bibitem{avi1}
S.~Hooy, N.~Linial and A.~Wigderson,
{\em Expander graphs and their applications},
Bull.\ Amer.\ Math.\ Soc. {\bf 46} (2006).

   \bibitem{avi2}
A.~Lubotzky
{\em Expander Graphs in Pure and Applied Mathematics},
Bull.\ Amer.\ Math.\ Soc. {\bf 49} (2012).

\bibitem{us3}
  J.~.L.~F.~Barbon and J.~M.~Magan,
  {\em Fast Scramblers, Horizons and Expander Graphs},
  JHEP {\bf 1208} (2012) 016,
  [arXiv:1204.6435 [hep-th]].





  \end{thebibliography}
\end{document}